\documentstyle[aps,epsf,prb,multicol]{revtex}
\newlength{\colwi}
\begin{document}
\raggedbottom
\twocolumn[\hsize\textwidth\columnwidth\hsize\csname@twocolumnfalse\endcsname
\draft
\preprint{\today}
\title{Cantori and dynamical localization in the Bunimovich Stadium}

\author{Fausto Borgonovi$^{[a,b,c]}$, Paolo Conti $^{[d]}$  
Daniela Rebuzzi$^{[d]}$, Bambi Hu$^{[e,f]}$, and Baowen Li$^{[e]}$  }
\address{
$^{[a]}$Dipartimento di Matematica e Fisica,
Universit\`a Cattolica,
via Trieste 17, 25121 Brescia, Italy  \\
$^{[b]}$Istituto Nazionale di Fisica Nucleare, Sezione di Pavia,
via Bassi 6, 27100 Pavia, Italy\\
$^{[c]}$Istituto Nazionale di Fisica della Materia, Unit\`a di Milano,
via Celoria 16, 22100 Milano, Italy\\
$^{[d]}$  Dipartimento di Fisica Nucleare e Teorica,
Universit\'a di Pavia, via Bassi 6, 27100 Pavia, Italy\\
$^{[e]}$ Department of Physics and Centre for Nonlinear Studies,
Hong Kong Baptist University, Hong Kong, China\\
$^{[f]}$ Department of Physics, University of Houston, TX77204, USA 
}

\maketitle
\begin{abstract}
Classical and quantum properties of the Bunimovich stadium in 
the diffusive regime are reviewed. In particular, the quantum
properties are directly investigated using an approximate quantum map.
Different localized 
regimes are found, namely, perturbative, quasi-integrable
(due to classical Cantori), dynamical  and ergodic. 
\end{abstract}
\pacs{PACS numbers:
05.45+b}]
 
\section{Introduction}
\label{sec:level1}

Long time ago, some physicists in the Siberian  
winter were playing dice with chaos
and quantum mechanics. This game 
resulted in a  paper which 
has been quoted N+1 times in literature\cite{ccfi} and  
which is known 
as a milestone in Quantum Chaos. 

Far from being an isolated branch of physics, Quantum Chaos has
revealed its importance in the last twenty years in many 
important physical applications of different fields: Solid State,
Nuclear, Atomic and Mesoscopic physics, just to give few
examples.

The common paradigma in Quantum Chaos is the so--called Kicked Rotator 
Model (KRM) whose   dynamics is described by the Chirikov
standard map (CSM).
It represents the safe retreat of many researchers working in Quantum
Chaos. Indeed, even if still now papers dealing with some hidden 
property of the Chirikov Standard Map appear\cite{balescu},
(after the monumental
work of Boris Chirikov published in 1979 \cite{boris}), 
or regarding new mathematical advances in the knowledge 
of the Floquet spectrum of the KRM\cite{dbforni},
we may 
say that its general behavior is quite well understood, at least
from the physical point of view.

In this paper we show how a 
different physical problem can be explained using  
old results borrowed from KRM. 
 
The model under current investigation is the Bunimovich 
Stadium\cite{buni}.
The properties of this model are well known
in literature, both from  physical and mathematical point
of view. Here we are interested in this particular billiard,
characterized by a straight line $2a$ much smaller than the semicircle
radius $R$ : 
$\epsilon = a/R \ll 1$. Preliminary   studies  
\cite{bcl} have shown that the classical motion in 
the angular momentum is diffusive and it can be conveniently 
described by a 2--d area preserving map.
In the same paper\cite{bcl} the study of the nearest neighbor
level spacing distribution (NNLSD) exhibits 
a different behavior depending on the energy range 
where the statistics is taken.
In particular, at fixed $\epsilon \ll 1$ the NNLSD shows
a smooth crossover from a Poisson to a Wigner-Dyson distribution 
as the energy is increased. The borders below and above 
which one can expect a particular distribution were theoretically
predicted and confirmed by numerical data\cite{bcl}. 
The region characterized by intermediate statistics\cite{felixrep}
can be  associated,
on the basis of a well defined picture\cite{felixrep,boca},
 with the presence of dynamical localization.
In a sense, the qualified adjective ``dynamical'' was, at that stage
quite inappropriate. 
Indeed only and indirect proof of dynamical localization were given, 
based on level statistics. The first example of localized eigenstates,
in the angular momentum basis, was given in \cite{dimafram}
for a rough deformation of a circular billiard. In this case, due to 
the finite number of harmonics describing the smooth modification 
of the boundary, direct exponential localization was found 
and the equality between 
quantum localization length and classical diffusion rate
established.

In this paper we enforce this viewpoint by studying directly the
quantum dynamics instead of  eigenfunctions and eigenvalues.
This can be efficiently realized, 
from the numerical point of view, only quantizing the classical 
map.
The obtained numerical data\cite{borgo}  indicate that   quantum
equilibrium distributions, differently from the  
classical ones, are algebraically localized 
in the angular momentum space.
Further numerical investigations\cite{cp} of the stadium 
eigenfunctions confirmed their algebraic localization in the in
angular momentum space.
Nevertheless it is possible to distinguish  among different 
quantum regimes.

Besides the perturbative regime, characterized by trivial 
periodic dynamical behavior, other different kind of
localization can be identified. The first one, called quasi--integrable,
has been associated to classical Cantori. Another one is marked 
by dynamical localization, despite the algebraic localized
distribution (with this word we mean a situation in which classical
diffusion rate and quantum localization length have the same numerical 
value).  
Recent analytical studies\cite{prange} support this analysis.
 
The paper is organized as the following: in section II we consider
the classical dynamics inside the Bunimovich stadium as given by
a suitable perturbed twist map.
Classical properties of the discontinuous map are then investigated 
in section III, while section IV is devoted to its quantum 
properties. Finally in section V the estimates about different
borders in the energy--parameter plane are summarized.

\section{Mapping the Bunimovich Stadium}
\label{sec:level2}
 
The ensemble dynamics of classical point particles having 
energy $E$, unit mass and 
initial angular momentum $l_0$, colliding elastically 
inside the Bunimovich Stadium is described, when  
$\epsilon \ll 1$, by the following map ($R=1$)\cite{bcl}

\begin{eqnarray} 
\bar{l} &=& l -2\epsilon\sin\theta \ 
 {\rm sgn } (\cos\theta)\sqrt{2E-l_0^2} \nonumber \\
\bar{\theta} &=& \theta +\pi -2 \ {\rm asin} (\bar{l}/\sqrt{2E})
\qquad {\rm mod-}2\pi
\label{mapc}
\end{eqnarray}

In (\ref{mapc}) $\theta$ is the angle measured from the center 
of the Stadium, $l$ the angular momentum measured from 
the same center and the overlined variables ($\bar{\theta}, \bar{l}$)
indicate the values taken after the collision with border. 
This map has been obtained by neglecting collisions
with the straight lines, terms $O(\epsilon^2)$ and
in the local approximation (small variations 
in the angular momentum).
 Indeed, it is easy to understand that while in the 
first equation of (\ref{mapc}) $l$ can grow infinitely,
the second equation  loses its validity
when $\vert l \vert $ approaches its maximum value $\sqrt{2E}$.  
The existence of such a bound is a trivial 
consequence of the energy conservation.
 
If we now put $l_0 = 0$ and 
approximate asin$(x)$ with its argument (since we must
exclude the values $\vert x \vert \simeq 1$), 
we get:
 
\begin{eqnarray} 
\bar{l} &=& l +k\sin\theta\ {\rm sgn}(\cos\theta) \nonumber \\
\bar{\theta} &=& \theta +T\bar{l} \qquad {\rm mod-}2\pi
\label{mapc1}
\end{eqnarray}

where we put 
$k = 2\epsilon\sqrt{2E} $, $T = 2/\sqrt{2E}$, $l \to -l$
and  $\pi$ 
 has been neglected since $\sin \theta \ {\rm sgn}(\cos\theta)$
is $\pi$-periodic. 
Map (\ref{mapc}) has been obtained\cite{bcl} for $\epsilon \ll 1$,
therefore   (\ref{mapc1}) holds when  
$kT = 4 \epsilon < 1$. The case  $kT > 1$, which represents
a possible regime for (\ref{mapc1}) has no physical meaning here
and it will not be taken into account.

Map (\ref{mapc1}) on the cylinder
$[0,2\pi) \times (-\infty,\infty)$ has been recently investigated
\cite{borgo} for $kT < 1$. Results can be then extrapolated to our case
assuming  $\vert l \vert \ll \sqrt{2E}$.
In the next sections we present a detailed study of classical and 
quantum properties of map (\ref{mapc1}).

\section{Discontinuous Twist maps}
\label{sec:level3}

Let us write (\ref{mapc1}) by introducing the variables 
$J = lT$, $K=kT$ in the following way :
\begin{eqnarray}
\bar{J} &=& J+ K \sin\theta\ {\rm sgn}(\cos\theta) \nonumber \\
\bar{\theta} &=& \theta +\bar{J} \qquad {\rm mod-}2\pi
\label{mapc2}
\end{eqnarray}

such that we have single parameter $K$.
Map (\ref{mapc2}) belongs to a particular class of discontinuous
twist maps.
We use the word "discontinuous" to mark the difference with
the Chirikov Standard Map, for which $\bar{J}-J= K\sin\theta$
is a continuous function in the interval $[0, 2\pi)$.

\begin{figure}
\epsfxsize 8cm
\epsfbox{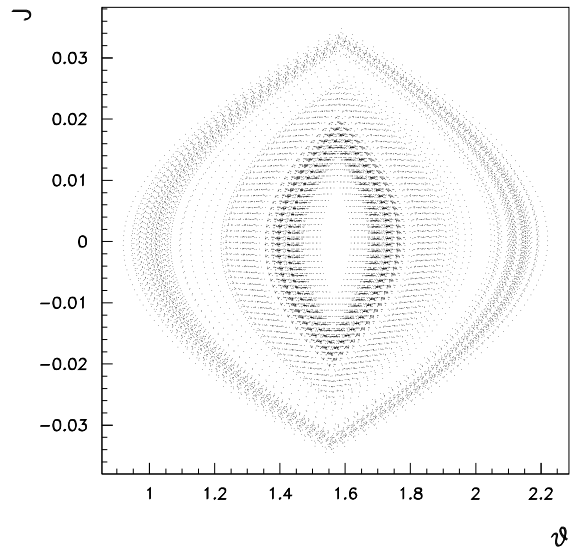}
\narrowtext
\caption{Poincar\`e surface of section for map (3)
and
$K=0.001$.
One particle
has been iterated $2\cdot 10^4$ times.
}
\label{poi}
\end{figure}

\begin{figure}
\epsfxsize 8cm
\epsfbox{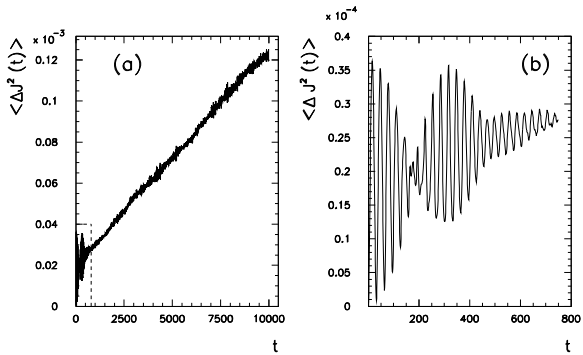}
\narrowtext
\caption{ Growth of the 
average momentum spreading in time 
for the discontinuous map (3) 
at $K=0.001$. b) is a 
magnification of the small area indicated among dashed
lines in the left corner of a).
The initial ensemble
consists of 2000 particles having the same momentum $J_0=0.1$ and
random phases $\theta$.
}
\label{mvt}
\end{figure}

The most studied case in the set of discontinuous
functions is the saw--tooth map (STM)
where the change in the angular  momentum is given by  
$\bar{J}-J= K(\theta-\pi)/\pi$.
 For this map the classical transport properties have been studied,
quite long ago\cite{dana}. 
The relevant difference of the STM, if compared with the CSM,
is characterized by the absence 
of a KAM structure. Indeed, since the hypothesis of KAM theorem
are not satisfied, we do not expect KAM tori exist for any 
$K$ value, independently from its smallness. 
Namely, differently from the CSM, where for $K<1$ the motion
is typically regular on invariant tori, here one finds
absence of KAM tori  for any $K$; nevertheless,
invariant structures still exist. Indeed Cantori can be defined
in the 
same way as in the STM\cite{dana}. 
An example of the motion in the neighborhood of Cantori 
is shown in Fig.\ref{poi}.
As one can see
the motion is far from being chaotic even if,
upon the increasing of  time, 
a single orbit can explore, in a dense way, the whole phase space. 
Moreover, the resulting motion is conveniently
described by a diffusive 
equation for the distribution function as can be inferred by
looking at 
the behavior in time of the average squared momentum (see Fig.\ref{mvt}a).

The linear growth in time, after a transient time
(see Fig.\ref{mvt})
and the corresponding Gaussian distribution in 
angular momentum, at a given time, (see for instance
\cite{bcl})
are usually taken as a common reference for the existence of
a diffusive motion\cite{boris}.

For such discontinuous maps, 
the resulting coefficient diffusion,
(extracted numerically
from a linear fit of Fig.\ref{mvt}a), 
can be shown to depend  
on the parameter $K$ in the following way\cite{dana,bcl,borgo}:

\begin{equation}
{\cal D} = \lim_{t\to\infty} \langle \Delta J^2 (t)  \rangle / t  
\propto K^{5/2}
\label{diff}
\end{equation} 

where $t$ is the iteration time. 

\begin{figure}
\hspace{.8cm}
\epsfxsize 8cm
\epsfbox{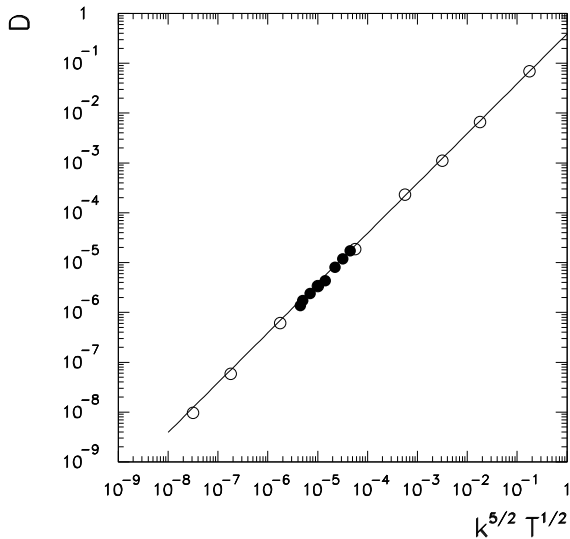}
\narrowtext
\caption{
Diffusion rate for the discontinuous map (2) for different $k$ and $T$
versus  $k^{5/2} T^{1/2}$.
Open circles are for $T=1$ , full circles for $k=0.01$
; both for $k T < 1 $. Full line is the best fit line $D=0.39 k^{5/2} T^{1/2}$.
}
\label{dsca}
\end{figure}

In terms of the original variables
one get   
$D = \langle \Delta l^2 (t)  \rangle / t \propto  k^{5/2} \sqrt{T}$
(see Fig.\ref{dsca}). 
The regime characterized by a diffusion coefficient scaling as $K^{5/2}$
has been called slow diffusion\cite{borgo}. This mark 
the difference 
from the standard quasilinear regime ($K > 1$) where typically one has
${\cal D} \simeq K^2$ (superimposed to oscillations\cite{reche}).

The dependence on the power $5/2$ has been explained\cite{dana}
in terms of a transport model based on a Markovian partition of the phase
space for the STM and 
it clearly indicates that the random  phase approximation\cite{boris} cannot
be applied in this case. 

Phases are indeed correlated within a time $\tau$
(see Fig.\ref{mvt} b)
during which the resulting 
motion cannot be chaotic nor diffusive.  
A correct evaluation of this time scale could be a key in 
understanding  the ``strange'' exponent $5/2$,
even if numerically it is quite difficult to obtain 
sharp results. 

\section{Quantum map dynamics}
\label{sec:level4}

Now we turn to the quantum analysis. Adopting the nowadays standard 
procedure\cite{ccfi} we analyze the quantum evolution of map (\ref{mapc1})
 by means 
of the one--period evolution operator 
given explicitly by  ($ \hbar =1 $):
\begin{equation}
{ \cal U}_T  = e^{-i T \hat{ n }^2 /2 } e^{-i k \vert \cos \theta\vert}
\label{qe}
\end{equation}

where $\hat{n} = -i\partial/\partial\theta$. 
Apart from the modulus in the potential $V(\theta)=\vert \cos\theta \vert $
the evolution operator $\cal{U}_T$ is exactly the same as the
Kicked Rotator one. 
Anyway the presence of the modulus leads to many important 
physical differencies. 
Indeed, written in the momentum basis $n$, the matrix
elements ${\cal U}_{n,m}= \langle n \vert {\cal U}_T  \vert m\rangle$
decay as a power law 
$\vert {\cal U}_{n,m} \vert \sim 1/\vert n - m \vert^2$
away from the principal diagonal (and not faster than 
exponentially as for the KRM). 
This case has been investigated for banded random matrices
\cite{fyo}, where it was shown that eigenfunctions are also algebraically
localized with the same exponent.

The presence of power law localized eigenstates has major consequences.
First of all it is not, a priori, obvious if the mechanism connected 
with the exponential dynamical localization holds even in this case.
Moreover, while in case of exponential localization a unique
measure of localization is defined
(up to a constant), for algebraical localization
different definitions of 
localizations can, in principle,
 give rise to different parametric dependences. 

Here we consider, as a measure of the degree of localization, the
variance $\xi_\sigma$ of the stationary  
distribution $P(n) = \vert \psi_n (t\to\infty) \vert^2$:

\begin{equation}
\xi_\sigma = [ \sum_n n^2 P(n) ]^{1/2}
\end{equation}

which we expect to have a well defined classical limit.

\subsection{The Kicked Rotator Model for  $kT<1$}

Before we turn to numerical results, it is useful to consider, as 
a common reference, the KRM. 
This is far from being pedagogical, since the case we are interested in here
($kT < 1$) has not been an object of intense investigations in the past.
Moreover, in this region, characterized by classical regular motion,
 it is quite difficult to get homogeneous results.
For instance, one should expect that the 
localization
length of the stationary distribution strongly depend on the initial 
conditions.
 Starting within a regular region (stable island
around periodic orbits) would result in a spreading width surely not 
larger than the size of the island (excluding the exponentially small
tunneling among different classical tori). This was indeed 
what Shepelyansky\cite{dima83} found: the spreading of the quantum 
stationary distribution can be roughly identified 
with the width $\simeq \sqrt{k/T}$ of
the main classical 
resonance, whose size is $\sqrt{k/T}$, when $kT < 1$. 
Another result was obtained  again by Shepelyansky by investigating
directly the quasi--energy 
eigenfunctions\cite{dima87} : he found 
direct proportionality between their localization lengths
and the parameter $k$ : $\ell \simeq k/4$.               
The different regimes in this undercritical case were also
reported by Izrailev\cite{felixrep}.
According to his Fig.3 (see also text) the case $K = kT < 1$ is marked by    
two different quantum 
borders.  One is the condition for the applicability of common
perturbative theory ($k\simeq 1$) while  the other one is the 
condition for the semiclassical
approach to describe quasiperiodic or chaotic motion (Shuryak border 
$k\simeq T$).
When $k<T$ the size of the nonlinear resonance is less 
than the distance between neighboring unperturbed levels and the 
quasiperiodic classical behavior is suppressed by quantum effects.

Our data on the behavior of 
$\xi_\sigma$ confirm 
and extend
this general 
picture. They can be summarized as follows:

1) $\xi_\sigma$ depends, for $K = kT < 1$,  only on the scaling parameter 
$k/T$.

2) as a function of $k/T$  two different regimes can be 
numerically detected, one linear, when $k < T$ (below the Shuryak
border) and another, for $k > T $, where $\xi_\sigma \propto \sqrt{ k/ T }$.

\begin{figure}
\epsfxsize 8cm
\epsfbox{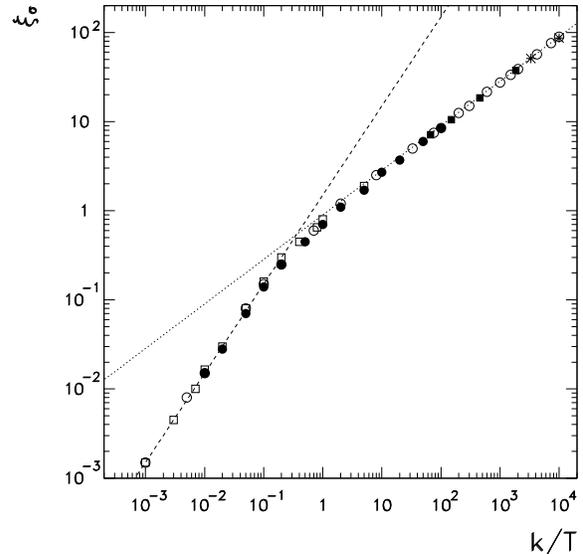}
\narrowtext
\caption{Localization length as a  function of the scaling
parameter $k/T$ for the Kicked Rotator and $kT < 1$.
Full symbols are obtained by fixing $k$ and varying $T$ :
circles ($k=0.01$), squares ($k=1$), asterisks ($k=10$).
Open symbols are obtained by fixing $T$ and varying $k$:
circles ($T=0.01$), squares ($T=0.1$).
Dashed line is $\  1.5 k/T$ (linear regime), while dotted line is
$0.9 \sqrt{k/T}$.
}
\label{lockr}
\end{figure}

These  results are presented in 
Fig.\ref{lockr}.

As one can see the scaling law is accurate up to eight orders
of magnitude.  The intersection point
between the two lines ( $k \simeq T $  called Shuryak border),
is the only transition point (for $\xi_\sigma$)
 we are able to detect
numerically. Results have been checked to be independent from the choice
of the initial state $\psi_n = \delta_{n,n_0}$. Since typically the variance
$\xi_\sigma (t) = \sqrt{ \langle \Delta n^2 (t) \rangle}$ is an oscillatory 
function of the iteration time $t$ (quasi--periodic motion)  the average
(in time) value has been taken. 
Fluctuations are typically quite large, often on the same order of the 
average value. 
On the basis of  previous results we identify two different
regimes, for the KRM in the classically 
regular case $kT<1$ :  the
perturbative one $\xi_\sigma \simeq k/T$ and the quasi--integrable one
$\xi_\sigma \simeq \sqrt{k/T}$. 

\subsection{The discontinuous quantum model in the slow diffusive case}

In this section we analyze the quantum behavior of the discontinuous 
map (\ref{mapc1}).
The classical  map is a good approximation to  the real billiard 
dynamics only for $\epsilon = kT/4 \ll  1$.
This means that we should consider, as a 
physical regime, only the slow diffusive one. 
The condition of applicability
$\vert l \vert /\sqrt{2E} < 1 $ will be considered
in details in section V.

\begin{figure}
\epsfxsize 9cm
\epsfbox{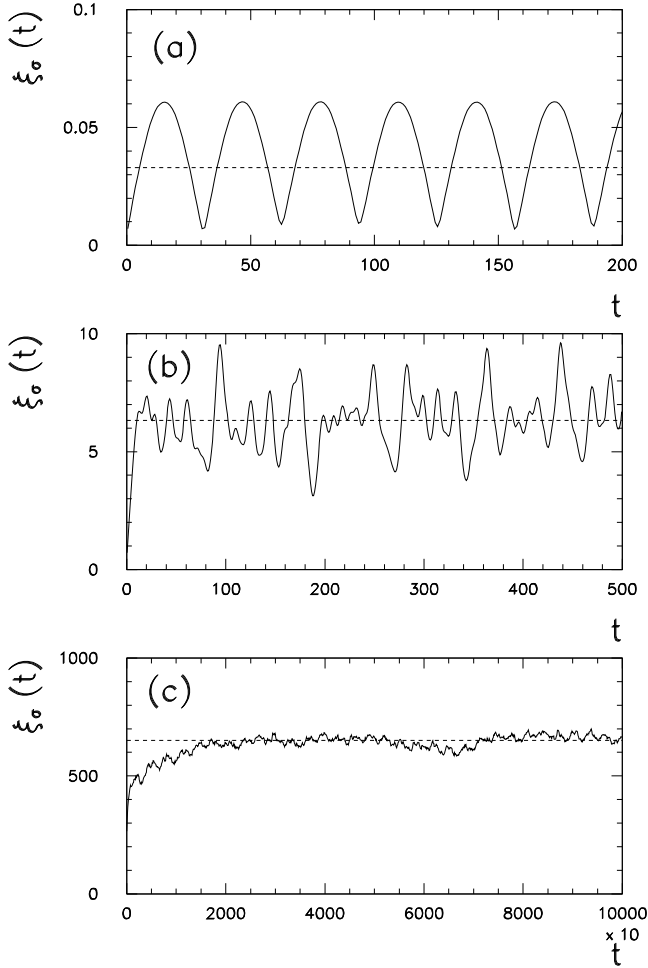}
\narrowtext
\caption{Wave packet spreading as a function of the iteration time $t$.
Here  $\xi_\sigma(t)=[ \sum_n n^2 \vert \psi_n (t) \vert^2 ]^{1/2}$
(a) is for $k=0.01$ and $T=0.1$ (perturbative region), dashed line
is the average $\xi_\sigma = 0.033$;
(b) is for $k=1$ and $T=0.01$ (quasi--integrable region), dashed line
is the average $\xi_\sigma = 6.33$;
(c) is for $k=100$ and $T=0.001$ (dynamical localization region), dashed line
is the average $\xi_\sigma = 650$.
}
\label{qn2}
\end{figure}

 Iteration of the quantum 
map (\ref{qe}) in this regime typically gives rise to 
the second moment $\langle \Delta n^2 (t) \rangle$ which is 
an oscillatory  function of the time $t$. 
In particular, it is possible to characterize values of the
parameters $k$ and $T$ which leads to periodic or irregular 
oscillations.
In  Fig.\ref{qn2} we show, for different $k$ and $T$, 
the behavior of $\xi_\sigma$  in time.
As dashed lines we indicate the average values over few oscillation 
periods. We will back to this picture later on when we discuss
the different quantum regimes.

\begin{figure}
\epsfxsize 8cm
\epsfbox{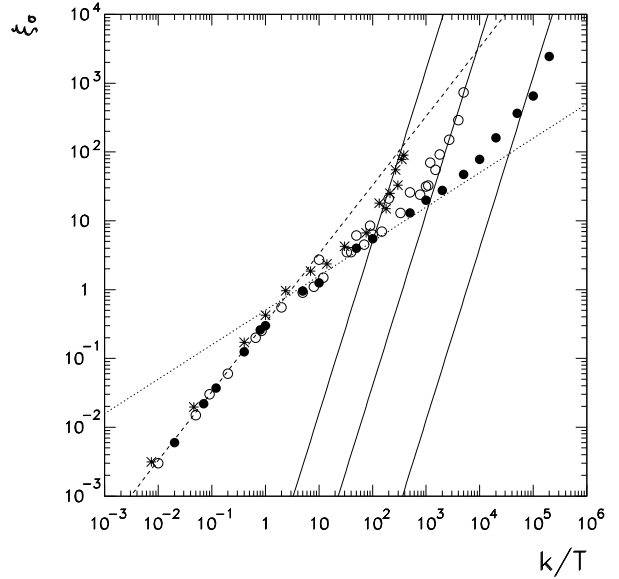}
\narrowtext
\caption{Localization length $\xi_\sigma$ as a function of $k/T$.
Three different sets of cases are shown, each one keeping $T$ fixed
and varying $k$: full circles ($T=0.001$), open circles ($T=0.01$)
and asterisks ($T=0.05$). Dashed line is $k/3T$ while dotted
is $\sqrt{k/T}/2$. Full lines are $\xi_\sigma = D(T)$ for the three
different sets, from the left to the right $T=0.05,0.01,0.001$.
}
\label{loc}
\end{figure}

By varying $k$ and $T$ and taking $\xi_\sigma$
as in Fig.\ref{qn2}, we obtain
three different scaling regions: the first
one for $k<T$ in which $\xi_\sigma \simeq k/T$, a second one for
$1/\sqrt{T} > k > T$ characterized by $\xi_\sigma \simeq \sqrt{k/T}$
and the third one, for $ 1/\sqrt{T} < k < 1/T$,
in which $\xi_\sigma \simeq D$ (see Fig.\ref{loc}).

It is easy to identify the first two regions in close analogy 
with those found for the KRM (see previous subsection).
While the existence of a perturbative region with the 
same characteristics of the KRM is not  surprising, much
care must be taken in interpreting the quasi--integrable
region. Indeed, while for the KRM we may properly speak about 
width of classical resonance, in this case we have no 
classical resonances at all. Nevertheless a close
inspection of Fig.\ref{poi} indicates the presence of
islands of ``quasi--integrability''  which means that trajectories
spend a lot of time before leaving from them. 
This region is indeed dominated by classical Cantori, which
act, from the quantum point of view as total barriers to the motion.
These effects, namely the quantum  propagation through 
classical Cantori have been investigated
in Ref.\cite{brown,geisel,macmei}. In particular, a  
relation between width of the holes of Cantori and $\hbar$
should exist, in order to obtain a meaningful semiclassical
limit. For instance Mackay and Meiss  proposed\cite{macmei} that 
Cantori could act as proper tori if the flux exchanged 
among different turnstiles is less than $\hbar$.

 Let us now analyze in more details the existence of a third 
scaling region for the localization length.
The classical analog of these   quantum regions of localizations
is a diffusive regime (after a transient time).
Nevertheless the localization found is not ``dynamical'' in the sense
that it cannot be derived for instance by taking the 
approach used 
by Chirikov, Izrailev and Shepelyansky  
for the KRM\cite{cis}. Moreover, since 
this kind of localization is shared
by the KRM, whose classical counterpart is regular motion,  
it cannot be followed by classical diffusive
excitation. A look at Fig.\ref{qn2} a-b confirms this view. 
 
 On the other side, if the typical relation of the dynamical 
localization $l_\sigma \simeq D$ holds true, the diffusion rate, in order
to produce an initial quantum classical-like diffusive spreading,
 has to be  
larger than the size of the classical ``quasi--resonance''.
This in turn  allows us to estimate the second transition point
as follows:

\begin{equation}
D = D_0 k^{5/2} \sqrt{T} =  c \sqrt{k/T}
\label{tp}
\end{equation}
 
which gives $k_{cr} = \sqrt{c/D_0 T}$ (where $c$ and $D_0$ are 
numerical constants of order one.
One can then guess that, if any, the dynamical localization regime 
can exist  for $k > k_{cr} = \sqrt{c/D_0 T}$.

\begin{figure}
\epsfxsize 9cm
\epsfbox{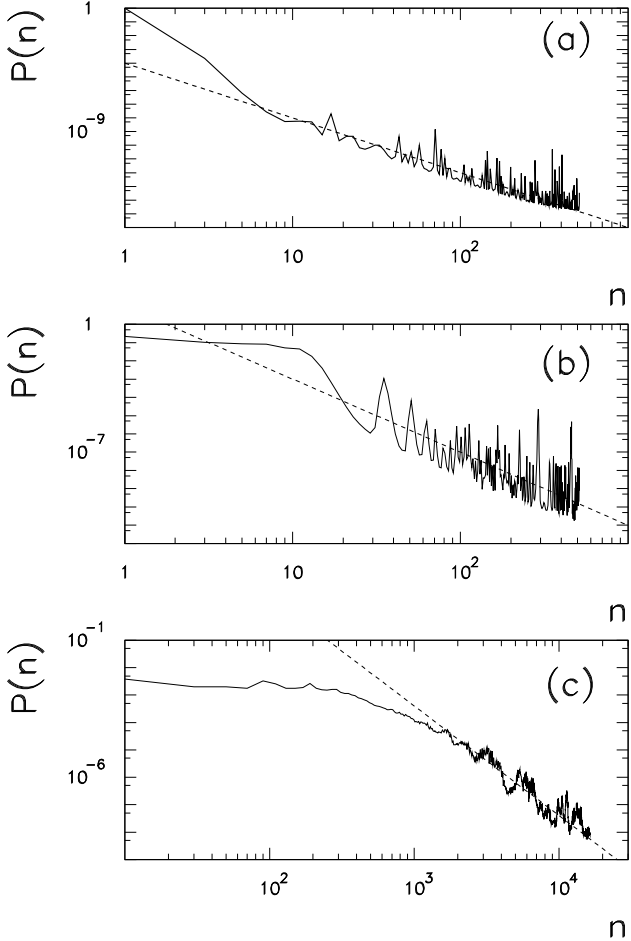}
\narrowtext
\caption{ Stationary distribution averaged over few oscillation
periods as a function of the momentum  $n$.
(a) is for $k=0.01$ and $T=0.1$ (perturbative region)
(b) is for $k=1$ and $T=0.01$ (quasi--integrable region)
(c) is for $k=100$ and $T=0.001$ (dynamical localization region).
Lines represent the power  $1/n^4$ and are shown to guide the eye.
}
\label{dis}
\end{figure}

Our numerical computations indicate that the dynamical localization
regime indeed exists, as can be inferred from Fig.\ref{loc}.
The sharp rise of the full lines shown in Fig.\ref{loc},
(one for each $T$ since the diffusion rate depends on $T$) 
indicates, without any doubt, the validity of the previous picture and
  the existence of the critical points, $k \simeq T$
and $k \simeq 1/\sqrt{T}$.

The existence of these thresholds for the Bunimovich stadium, 
and the regime of quasi--integrability as well, has
been  confirmed analytically by Prange, Narevich and Zaitsev\cite{prange}. 
Indeed they
were able to find an analytical expression for the eigenfunctions
up to  $\epsilon \sqrt{E} \simeq 1 $, which in terms of 
map variables  reads
$ k \simeq 1/\sqrt{T}$. 
Above the threshold $\epsilon \sqrt{E} \simeq 1$ the
semiclassical perturbative approach fails. According to our point of view 
this represent the $k$ value 
necessary to 
start the classical-like diffusion process in presence  of slow 
diffusion. 

Let us add few comments about the shape of the stationary distribution.
In all cases we have found good agreement with a power law distribution
 $P(n) \simeq \vert n-n_0 \vert^{-4}$. 
In Fig.\ref{dis}  we show the correspondent stationary distributions
for the cases of Fig.\ref{qn2}.  The lines, indicating the power law
behavior, are drawn to guide the eye.
This means that results obtained from banded random matrices\cite{fyo}
 can 
be extrapolated even 
when both randomness and 
dynamical chaos are absent (perturbative and 
quasi--integrable regions).

The presence of power law localized states for the discontinuous 
map, also recently confirmed in \cite{cp} for the stadium 
eigenfunctions, should be somehow put in relation with the exponential 
localization found in \cite{dimafram} for the eigenfunctions
of a rough billiard. This peculiarity should be in turn related
with the particular boundary shape perturbation. Indeed, while in 
the rough billiard considered in \cite{dimafram}, a finite number $M$ of 
harminics is necessary in order
to produce the deformation from the circle, the stadium boundary
perturbation is not analytical. The classical 
rough map is shown\cite{dimafram}
 to be chaotic when $\epsilon > \epsilon_c \sim M^{-5/2}$.
It is then clear that when $M\to \infty$ KAM regular structures disappear
and the situation depicted by the discountinuous map appears.

\begin{figure}
\epsfxsize 8cm
\epsfbox{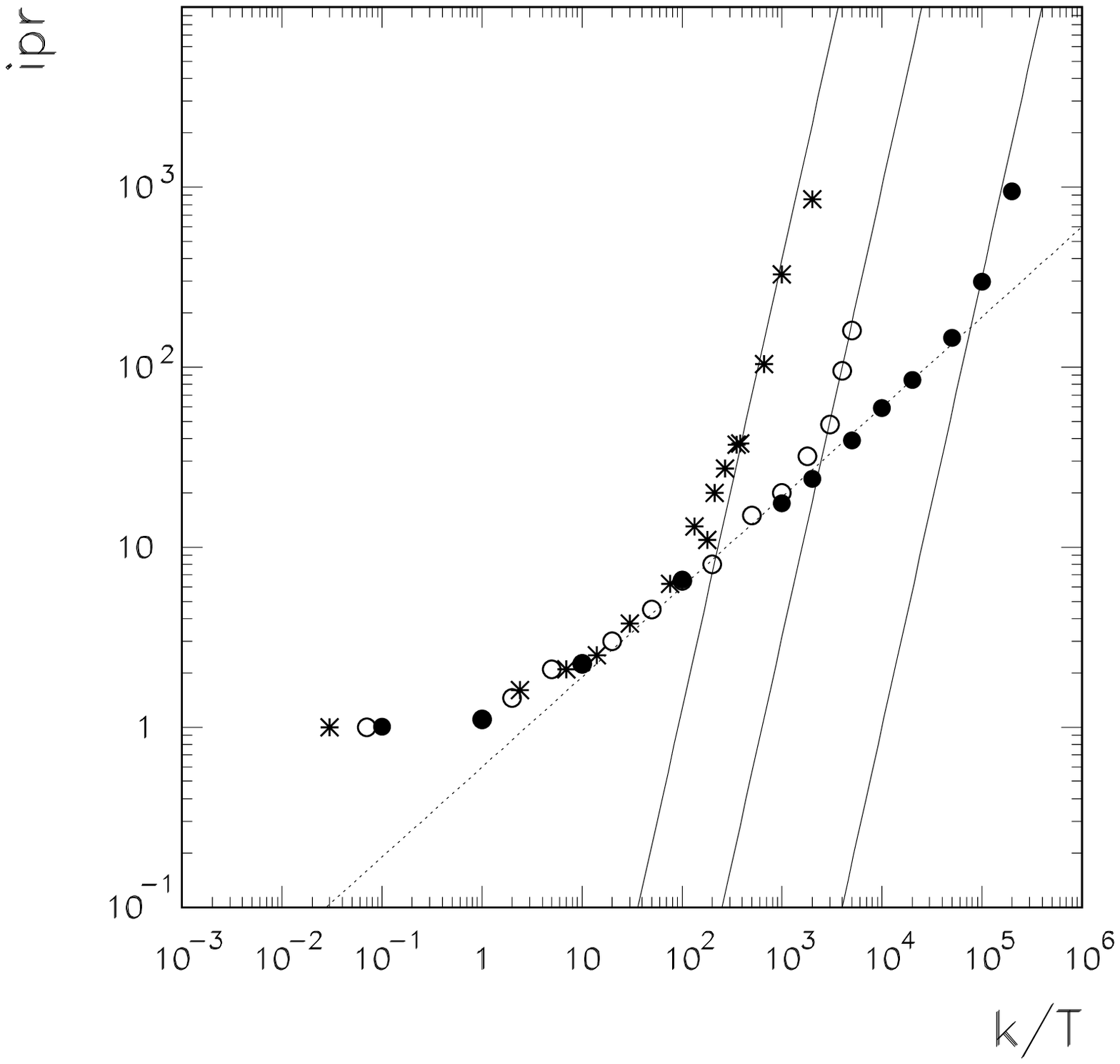}
\narrowtext
\caption{Inverse participation ratio $ipr$ as a function of $k/T$.
Three different sets of cases are shown, each one keeping $T$ fixed
and varying $k$: full circles ($T=0.001$), open circles ($T=0.01$)
and asterisks ($T=0.05$). Dotted line is $0.6 \sqrt{k/T}$.
Full lines are $ipr = D(T)/4$ for the three 
different sets, from the left to the right $T=0.05,0.01,0.001$.
}
\label{ipr}
\end{figure}

To end this section let us remark that a similar behavior happens
if another definition of localization length is taken.
In  Fig.\ref{ipr} 
we show the inverse participation ratio
of the quantum distribution  $ipr = 1/\sum_n \vert \psi_n \vert^4$
as a function of the scaling parameter $k/T$.
The absence of a perturbative region is due to the fact that, 
by definition, $ipr \ge 1$. Moreover, let us note that a numerical
constant has to be added in order to follow numerical data. 
Namely, the dynamical localization regime is marked by $ipr \simeq D/4$.
The simple proportionality between $l_\sigma$ and $ipr$,
even if not surprising (the same happens for the KRM), has
to be considered accidental for power law localized distributions.

\section{Borders for the Stadium}

The results found in the previous sections can be 
extended to the Bunimovich Stadium. This will lead   
to important estimates which enable us to discriminate between different
physical situations. In particular  the   critical 
points found previously  give rise to relations between the energy 
and the small parameter $\epsilon$.

\begin{figure}
\hspace{.8cm}
\epsfxsize 6cm
\epsfbox{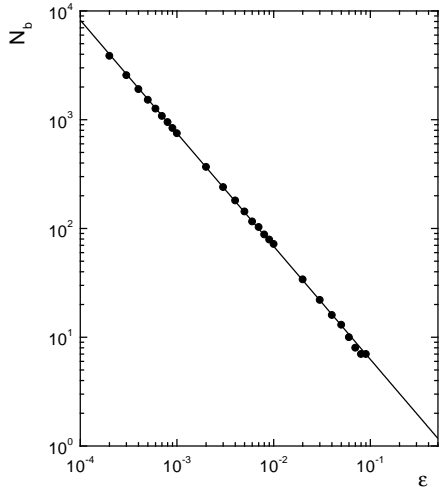}
\narrowtext
\caption{Perturbative border for the Bunimovich Stadium.
$N_b$ is the quantum number corresponding to the eigen energy
$E$. Solid circles
represent numerical data, and straight line is the best-fit
 $N_b = 0.57/ \epsilon^{1.04}$.}
\label{pb} \end{figure}

The first regime (perturbative), 
is characterized by  $ k < T$ or $2 E\epsilon < 1$. 
From the billiard point of view this means that  levels having 
an energy less than $E \sim 1/\epsilon $ can be obtained 
perturbatively from those of the unperturbed spectrum ($\epsilon=0$, e.g.
the circle). 
The numerical study of the energy spectrum gives a 
confirmation for this border.
We label the energy levels as $E_N(\epsilon)$. If $\epsilon$ is
sufficiently small we can assume the shift 
$\Delta E = E_N(\epsilon) - E_N (0)$ to be quite small.
On the other hand the energy spectrum is characterized by an average level
spacing $\delta E_N$ which is given 
approximately by the Thomas-Fermi formula\cite{bohigas}:

\begin{equation}
\delta E_N \simeq {{E_N}\over{N_b}} \simeq { {8\hbar^2}\over {m R^2}} 
\label{tf}
\end{equation}

where $m=1$ is the particle mass, $R=1$ is the circle radius
and $N_b$ is the number of levels up to energy $E$.
In order to be in a perturbative
regime there should be 
no levels overlapping, namely 
$\Delta E_N(\epsilon) < \delta E_N$. 
This gives a relation between the energy $E$
(or the level number $N_b$)  and the small parameter $\epsilon$,
which can be detected numerically. We show our numerical results in 
Fig.\ref{pb}. The best fit gives rise to $0.57/\epsilon^{1.04}$, which is 
in a good agreement with the theoretical prediction.
Numerically, $N_b$ is obtained by comparing the eigenernergy list of 
the circular billiard and that of the perturbed Bunimovich stadium 
of $\epsilon$.

Moreover, a kind of semiclassical approach
is possible\cite{prange} up to $k = 1/\sqrt{T}$, or
$E \sim 1/\epsilon^4$ (we omit a numerical constant in front of
this expression : its value can be obtained  only numerically).
This means that the  analytical approach to eigenvalues and 
eigenvectors is possible up to this energy value. Nevertheless,
as soon as the map dynamics correctly
approximates  the real dynamics,
the  localization length 
should have a different energy dependence. 
Even in this case a direct numerical approach is needed in order 
to give a definite answer.

Let us now come to the much more interesting case marked 
by the dynamical localization. We have found that this
is possible only for $E >  1/\epsilon^4$. According to the
dynamical localization theory the quantum spreading
will occur up to a time $t_B \simeq D $ called break time. 
Following \cite{bcl} we may speak of a proper localized
quantum regime only if $t_B < t_{erg}$, where 
$t_{erg}$ is the classical time in order to reach a stationary
ergodic distribution, estimated as 
$ \Delta l^2 \simeq D t_B $ or $ t_B \simeq \epsilon^{-5/2}$.
Would $t_B$ be larger than $t_{erg}$, classical and quantum
distributions will both reach an ergodic stationary distribution.
Putting $t_B = t_{erg}$ or $E \sim \epsilon^{-5}$
we get 
the last critical point 
 above which we expect 
quantum as well classical ergodicity.

The approach to ergodicity cannot be studied using this map. Indeed 
it represents a good approximation to real dynamics only for
$t \ll t_{erg}$.  This means that 
our results cannot be compared with those found by Frahm and 
Shepelyansky\cite{fs2} about  the approach to ergodicity via 
a Breit--Wigner regime. This is a situation characterized by
eigenfunctions delocalized on the energy shell but with many 
strong isolated peaks of probability. The presence of isolated 
peaks of probability in our case (see Fig.7) is instead due 
to the classical phase space structure.

While the ergodicity regime has been previously investigated\cite{bcl}
using the NNLSD, the existence of two different localized regimes
($\epsilon^{-1} < E < \epsilon^{-4}$ and $\epsilon^{-4} < E < \epsilon^{-5}$)
has been only guessed on the basis of the similarity with the 
approximate map. 
In this case, the study of NNLSD should be probably 
accomplished  by a direct study of eigenfunctions.
This we argue, since it is not at all obvious 
how different kind of localization can affect 
the level statistics.
A preliminary study in this direction can be found in \cite{cp}.

Let us note that, in order to correctly approximate real dynamics
with the map, one has to require  $\vert l\vert/\sqrt{2E} < 1 $,
or $lT<2$. One should then require  that in both,
quasi--integrable and dynamical localization regime,
$\xi_\sigma T < 2 $.
This in turn means either
\begin{equation}
\sqrt{{k\over T}} T = \sqrt{kT} < 2
\end{equation}
(which is always justified since $kT = 4\epsilon$ and $\epsilon < 1$) or
$$
 k^{5/2} T^{3/2} < 2
$$
which reduces to
$$
k (kT)^{3/2} \simeq
 \epsilon E^{1/2} \epsilon^{3/2} = E^{1/2} \epsilon^{5/2} < 2
$$
or $E < 1/\epsilon^5$ which is the condition 
in order to have classical ergodicity.
This is the reason why we neglect this condition in Sec. IV.

Most of the results were obtained by approximating the real
quantum dynamics by means of a quantum map which is
the quantum analog of a classical map approximating
the classical dynamics. This kind of procedure is not new
(see for instance \cite{hatom}). One may wonder if this
is, at the end, close to the original model. The answer
can come, of course, only  from a direct numerical or experimental,
analysis of the quantum dynamics of  wave packets inside the billiard.
Nevertheless it is significative that different approaches 
on the same model\cite{prange} 
give results in good agreement with ours.

After the completion of this work we became aware of other related 
works on the subject\cite{cp}. In particular their numerical data,
while confirming the existence of these regimes, 
indicate other different borders. However, from the map 
point of view, the transition at $k=T$ ($E \sim 1/\epsilon$)
 is very sharp, we cannot exclude
numerically the presence of a further border at 
$k \sim T^{1/3}$ ($E\sim \epsilon^{-3}$).
Indeed a close inspection at Fig.6, indicates  a very smooth transition 
toward the line $\xi_\sigma = D$. Further numerical calculations are required
in order to show if quantum map also shows this border.

\section{Acknowledgments}
We thank R. Prange, R.Narevich  and O.Zaitsev for 
making their work available before publication. Discussions
with G.Casati are also acknowledged. BH and BL were supported in part by 
the grants from the Hong Kong Research Grants Council (RGC) and the Hong 
Kong Baptist University Faculty Research Grants (FRG).

\end{document}